\documentclass[aps,preprint,prd,floatfix,nofootinbib]{revtex4-1}
\usepackage{amsfonts}
\usepackage{amsmath}
\usepackage{amssymb}
\usepackage[utf8]{inputenc}
\usepackage{float}
\usepackage{graphicx}%
\usepackage{color}
\providecommand{\U}[1]{\protect\rule{.1in}{.1in}}

\begin{document}
\title{Scalar field quasinormal modes on asymptotically locally flat rotating black holes in three dimensions}
\author{Andrés Anabalón$^{1,2}$, Octavio Fierro$^{3}$, José Figueroa$^{4}$ and Julio Oliva$^{4}$}
\address{
$^1$ Departamento de Ciencias, Facultad de Artes Liberales,
Universidad Adolfo Ibáñez, Avenida Padre Hurtado 750, Viña del Mar, Chile,\\
$^2$ Max-Planck-Institut für Gravitationsphysik, Albert-Einstein-Institut, 14476 Golm, Germany,\\
$^3$ Departamento de Matemática y Física Aplicadas, Universidad Católica de la Santísima Concepción, Alonso de Rivera 2850, Concepción, Chile\\
$^4$ Departamento de Física, Universidad de Concepción, Casilla, 160-C, Concepción, Chile.}
\email{andres.anabalon@uai.cl, ofierro@uscs.cl, josepfigueroa@udec.cl, juoliva@udec.cl}

\begin{abstract}
The pure quadratic term of New Massive Gravity in three dimensions admits
asymptotically locally flat, rotating black holes. These black holes are
characterized by their mass and angular momentum, as well as by a hair of gravitational origin. As in the Myers-Perry solution in dimensions greater than
five, there is no upper bound on the angular momentum. We show that, remarkably, the
equation for a massless scalar field on this background can be solved in an
analytic manner and that the quasinormal frequencies can be found in a closed form. The spectrum is obtained requiring ingoing boundary conditions at the horizon and an asymptotic behavior at spatial infinity that provides a well-defined action principle for the scalar probe. As the angular momentum of the black hole approaches zero, the imaginary part of the quasinormal frequencies tends to minus infinity, migrating to the north pole of the Riemann Sphere and providing infinitely damped modes of high frequency. We show that this is consistent with the fact that the static black hole within this family does not admit quasinormal modes for a massless scalar probe.
\end{abstract}
\maketitle

\section{Introduction}

Three dimensional gravity has been a useful arena to explore gravitational
models with simpler properties than their counterpart in four dimensions. Since the Weyl tensor identically vanishes in three dimensions,
Einstein equations imply that the spacetime is locally flat or (A)dS depending
on whether the theory has a null, negative or positive cosmological term,
respectively. This implies that General Relavitity has no local degrees of
freedom in vacuum, but nevertheless in the case with a negative cosmological
constant, its spectrum contains black holes (the BTZ black hole) \citep{BTZ}.
These black holes have been of fundamental importance on the tests of the
holographic correspondence between physics in Anti de Sitter spacetime and
that of a dual Conformal Field Theory living at the boundary \citep{Maldacena:1997re}, \citep{Witten:1998qj}, \citep{Gubser:1998bc}. Just to mention
two examples of this relation, the entropy of the BTZ black holes can be
obtained by a microscopic counting of microstates in the dual theory
\citep{Strominger} and their quasinormal ringing correlates
precisely with relaxation time in the dual field theory at finite temperature
\citep{Horowitz:1999jd}, \citep{Birmingham:2001pj}. By the end of the last decade it was also
realized that the lack of local degrees of freedom of General Relavitiy in
three dimensions allows to construct a parity invariant, self-interacting theory for a massive particle of spin 2, i.e., a massive
gravity theory. The Einstein-Hilbert Lagrangian is supplemented by quadratic
terms in the curvature, and despite the fact that the field equations are of
fourth order, their linearization around flat space correctly reproduces the
Fierz-Pauli equation for a massive spin 2 excitation \citep{NMG}-\citep{Deser:2009hb}. The addition of a cosmological term allows to
further explore the ideas of the holographic correspondence in the presence of
a massive graviton in the bulk. As it generically occurs in theories
containing quadratic powers in the curvature, such theory possesses two
maximally symmetric (and therefore of constant curvature) solutions. For a
precise relation of the couplings both vacua coincide and actually in this
case one can find asymptotically locally (A)dS black holes that are
characterized by the mass, angular momentum and an extra parameter that stands
for a gravitational hair \citep{OTT}-\citep{Bergshoeff:2009aq}. As shown in \citep{Giribet:2009qz} the entropy of such
black holes can be reproduced by counting microstates in the boundary theory,
providing a new test for such relation on a theory with local degrees of freedom
in the bulk. Recently, there has been a revived interest in the study of
asymptotic symmetries to null surfaces, where such surface could be null
infinity or the event horizon of a black hole. It is expected that
these studies might shed some light on the information paradox (see e.g.\citep{Arcioni:2003td}-\citep{Donnay:2015abr} and references therein). New Massive Gravity provides for a simple setup to carry on such
studies, since the spectrum of the purely quadratic theory (that is healthy and
intrinsically of fourth-order \citep{Deser:2009hb}) contains asymptotically locally
flat, rotating black holes in 2+1 dimensions \citep{OTT}. Such black holes do not
exist in General Relativity in vacuum. Within the realm of NMG, these black holes can be
generalized to construct non-circular black objects, dubbed black flowers
\citep{Barnich:2015dvt}, a family of metrics whose simplest representative is the
rotating black hole constructed in \citep{OTT} in the massless limit of NMG\footnote{It is also worth mentioning that these solutions can be embedded in the Born-Infeld extensions of New Massive Gravity \cite{BINMG} by removing the Einstein-Hilbert term \cite{BFINBINMG}.}. The aim
of this paper is to show that the equation for the massless scalar,
remarkably, can be solved analytically on these backgrounds, and that
the quasinormal frequencies can be found in a closed manner. As usual, we require
ingoing boundary condition at the horizon. We show that at infinity the
natural boundary condition that makes the action principle for the scalar to
be well-defined is that the field must vanish sufficiently fast. This is one
of the few known, rotating black holes, that admit such integration.

This paper is organized as follows: Section I introduces the theory of New
Massive gravity and presents the asymptotically locally flat, rotating black
hole solution. In Section II we consider a massless scalar field perturbation
and solve it in an exact and analytic manner. We find the quasinormal modes in a closed form in the rotating case and show that as the angular momentum of the black hole decreases the imaginary part of the frequencies approach minus infinity. This is consistent with the fact that in the static case, the algebraic equations that determine the spectrum cannot be fulfilled. We therefore explore the behavior of the
quasinormal frequencies in terms of the global charges of the background solution. Even though our interests is in the propagation of a scalar probe on the black hole background, let us mentions that our results could shed some light on the stability of the rotating black hole in NMG. Since the full non-linear field equations of NMG are of fourth order, their linearized version around a generic background will be in general of the same order. Nevertheless, there could be an affective quantity, constructed with second derivatives of the perturbation, which may have a second order dynamics, as it occurs in the Teukolsky equation, where the unknowns are linearized expressions for the components of the Weyl tensor in a null tetrad \cite{Chandrasekhar}. It is also worth mentioning that in some particular cases in GR it is known that the dynamics of some of the modes of the full gravitational perturbation coincide with the dynamics of a scalar probe with a given mass. This happens for example in the massless topological black hole \cite{Birmingham:2006zx}. A less striking connection between a scalar probe and a gravitational perturbation occurs with the Regge-Wheeler equation, which has the same functional form than the equation for a massless scalar probe in Schwarzschild background, but in this case there is a single numeric coefficient of difference in the effective Schroedinger potential (see \cite{Chandrasekhar}).  Section III contains conclusions and further comments.

\section{Massless limit of NMG and its rotating black hole}

The theory of New Massive Gravity \citep{NMG}, has the following action%
\begin{equation}
I\left[  g\right]  =\frac{1}{\kappa^{2}}\int d^{3}x\sqrt{-g}\left[
-R+\frac{1}{m^{2}}\left(  R_{\mu\nu}R^{\mu\nu}-\frac{3}{8}R^{2}\right)
\right]  \ .
\end{equation}
Note that the Einstein term has the wrong sign, which is necessary to obtain a
ghost free theory for a massive graviton, at the linearized level around flat
space. Since General Relativity does not propagate local degrees of freedom in
three dimensions, we can choose the sign of the Einstein term at convenience.
The massless limit is described by the purely quadratic theory which acquires
a linearized conformal invariance \citep{Bergshoeff:2009aq}, and despite the fact of having
fourth order field equations, it defines a healthy theory \citep{Deser:2009hb}. We take the
limit in the following manner $m\rightarrow0$ and $\kappa\rightarrow+\infty$
with $16\pi G:=\kappa^{2}m^{2}$ fixed. The field equations in vacuum are
therefore given by%
\begin{equation}
K_{\mu\nu}=2\square R_{\mu\nu}-\frac{1}{2}\left(  \nabla_{\mu}\nabla_{\nu
}R+g_{\mu\nu}\square R\right)  -8R_{\mu\rho}R_{\nu}^{\ \rho}+\frac{9}%
{2}RR_{\mu\nu}+g_{\mu\nu}\left(  3R^{\alpha\sigma}R_{\alpha\sigma}-\frac
{13}{8}R^{2}\right)  =0\ . \label{pureK}%
\end{equation}
Diffeomorphism invariance of the original action ensures the identity
$\nabla^{\mu}K_{\mu\nu}\equiv0$. This is the unique quadratic combination in
three dimensions that leads to field equations of fourth order, whose trace
reduces to a second order constraint \citep{Nakasone:2009bn}. This feature is of fundamental
importance when proving the healthiness of the theory. The theory defined by the field
equations (\ref{pureK}) has the following rotating black hole solution%
\begin{equation}
ds^{2}=-\left(  br-\mu\right)  dt^{2}+\frac{dr^{2}}{br-\mu}-a\left(
br-\mu\right)  dtd\phi+\left(  r^{2}+r_{0}^{2}\right)  d\phi^{2}\ ,
\label{therotatingblackhole}%
\end{equation}
with%
\begin{equation}
r_{0}^{2}=\frac{a^{4}b^{2}+16a^{2}\mu}{64}\ .
\end{equation}
It can be checked that this spacetime has a vanishing Cotton tensor, and it is
therefore a solution of Conformal Gravity in $2+1$ dimensions (the massless limit of Topologically Massive Gravity \citep{TMG1}, \citep{TMG2}). In such
context, the static solution was originally reported in \citep{Oliva:2009hz}. The
spacetime (\ref{therotatingblackhole}), is characterized by three parameters
$b$, $a$ and $\mu$ which determine the global charges associated to the
asymptotic Killing vectors $\partial_{t}$ and $\partial_{\phi}$, i.e. the mass
and the angular momentum, respectively, which are given by \citep{Barnich:2015dvt}%
\begin{equation}
M=\frac{b^{2}}{32G}\ ,\ J=Ma. \label{globalcharges}%
\end{equation}
Note that the parameter $\mu$ does not appear in the global charges, and
therefore can be interpreted as a hair of gravitational origin. Further
thermodynamical properties are the temperature and the entropy of the black
hole that respectively read
\begin{equation}
T=\frac{b}{4\pi},\text{ }S=\frac{\pi b}{4G}\ .
\end{equation}
The remaining chemical potential, the angular velocity of the horizon, vanishes.
The metric (\ref{therotatingblackhole}) can be obtained from the rotating
solution of Cosmological New Massive Gravity at the point of coinciding vacua
\citep{OTT}, in the limit $l\rightarrow+\infty$, by a suitable rescalling of the
integration constants. This solution can also be written in null coordinates,
adapted to future null infinity as%
\begin{align}
ds^{2}  &  =-\frac{\left(  8r+a^{2}b\right)  ^{2}\left(  br-\mu\right)
}{64r^{2}+16a^{2}\mu+a^{4}b^{2}}du^{2}-\frac{2\left(  8r+a^{2}b\right)
}{\sqrt{64r^{2}+16a^{2}\mu+a^{4}b^{2}}}dudr\label{solinnull}\\
&
\ \ \ \ \ \ \ \ \ \ \ \ \ \ \ \ \ \ \ \ \ \ \ \ \ \ \ \ \ \ \ \ \ \ \ \ \ \ \ \ \ \ \ \ \ \ \ +\left(
r^{2}+r_{0}^{2}\right)  \left(  d\psi-\frac{32a\left(  br-\mu\right)
}{64r^{2}+16a^{2}\mu+a^{4}b^{2}}du\right)  ^{2}\ ,
\end{align}
where the change of coordiantes that relates (\ref{therotatingblackhole}) and
(\ref{solinnull}) is%
\begin{align}
du  &  =dt-\frac{\sqrt{64r^{2}+16a^{2}\mu+a^{4}b^{2}}}{\left(  8r+a^{2}%
b\right)  \left(  br-\mu\right)  }dr\ ,\\
d\psi &  =d\phi-\frac{32a}{\left(  br-\mu\right)  \sqrt{64r^{2}+16a^{2}%
\mu+a^{4}b^{2}}}dr\ .
\end{align}

Hereafter we will use the metric in the coordinate system defined in equation
(\ref{therotatingblackhole}). The event horizon is located at $r=r_{+}=\mu/b$,
and the metric is asymptotically locally flat, i.e.%
\begin{equation}
\lim_{r\rightarrow+\infty}R_{\ \ \ \alpha\beta}^{\mu\nu}=0\ .
\end{equation}

\section{Massless scalar on the rotating black hole}

Consider a massless scalar $\Phi$, on the rotating black hole metric
(\ref{therotatingblackhole})%
\begin{equation}
\square\Phi=0\ .
\end{equation}
As usual, the symmetries of the background allow to consider the following
mode decomposition%
\begin{equation}
\Phi\left(  t,\phi,r\right)  =\Re\left(  \sum_{n=-\infty
}^{+\infty}\int d\omega\ e^{-i\omega t+in\phi}R_{\omega,n}\left(  r\right)
\right)  \ .
\end{equation}
Hereafter we denote $R_{\omega,n}\left(  r\right)  =R\left(  r\right)  $. One
therefore obtains an ODE for the radial dependence, which acquires a simple
fashion in terms of the coordinate $x$, such that%
\begin{equation}
r=\frac{x+\mu}{b}\ ,
\end{equation}
that maps the domain $r\in\lbrack r_{+},+\infty\lbrack$ to the range
$x\in\lbrack0,+\infty\lbrack$. The equation for $R\left(  x\right)  $
therefore reads%
\begin{equation}
A\left(  x\right)  \frac{d^{2}R\left(  x\right)  }{dx^{2}}+B\left(  x\right)
\frac{dR\left(  x\right)  }{dx}+C\left(  x\right)  R\left(  x\right)  =0\ ,
\label{ode}%
\end{equation}
with%
\begin{align}
A\left(  x\right)   &  =b^{2}x^{2}(a^{2}b^{2}+8\mu+8x)^{2}\ ,\\
B\left(  x\right)   &  =b^{2}x(a^{2}b^{2}+8\mu+16x)(a^{2}b^{2}+8\mu+8x)\ ,\\
C\left(  x\right)   &  =(a^{2}b^{2}(a^{2}b^{2}+16\mu)+64(x+\mu)^{2})\omega
^{2}-64b^{2}nx(a\omega+n)\ .
\end{align}

Near the event horizon, located at $x=0$, the solutions of equation (\ref{ode}) allow for the following asymptotic behaviors%
\begin{equation}
R\left(  x\right)  =C_{1}x^{-i\frac{\omega}{b}}\left(  1+\mathcal{O}\left(
x\right)  \right)  +C_{2}x^{i\frac{\omega}{b}}\left(  1+\mathcal{O}\left(
x\right)  \right)  \ ,
\end{equation}
while at infinity the scalar behaves as%
\begin{equation}
R\left(  x\right)  =\frac{\tilde{C}_{1}}{x^{\delta_{+}}}\left(  1+\mathcal{O}%
\left(  \frac{1}{x}\right)  \right)  +\frac{\tilde{C}_{2}}{x^{\delta_{-}}%
}\left(  1+\mathcal{O}\left(  \frac{1}{x}\right)  \right)  \ ,
\end{equation}
where%
\begin{equation}
\delta_{\pm}=\frac{1}{2}\left(  1\pm\sqrt{1-\frac{4\omega^{2}}{b^{2}}}\right)
\ .
\end{equation}
Here $C_{1,2}$ an $\tilde{C}_{1,2}$ are integration constants. Note that since
the angular velocity of the horizon vanishes, the leading dependence of the
scalar on the near horizon region does not depend on the rotation parameter
$a$.

A remarkable fact is that even though (\ref{ode}) describes the radial
dependence of a scalar on a rotating black hole,
it can be solved in an analytic manner in terms of Hypergeometric functions.
The solution of (\ref{ode}) is given by%
\begin{align}
R\left(  x\right)   &  =\left(  8x+8\mu+a^{2}b^{2}\right)  ^{-i\frac{\sqrt
{2}\left(  a\omega+2n\right)  }{\sqrt{a^{2}b^{2}+8\mu}}}x^{-i\frac{\omega}{b}%
}\left[  C_{1}F\left(  \alpha,\beta,\gamma,-\frac{8x}{a^{2}b^{2}+8\mu}\right)
\right. \nonumber\\
&  \left.  +C_{2}\left(  -\frac{8x}{a^{2}b^{2}+8\mu}\right)  ^{1-\gamma
}F\left(  \alpha+1-\gamma,\beta+1-\gamma,2-\gamma,-\frac{8x}{a^{2}b^{2}+8\mu
}\right)  \right]  \ , \label{lasolucion}%
\end{align}
where $F$ stands for a Hypergeometric Function $_{2}F_{1}$, and%
\begin{align}
\alpha &  =\frac{(b\delta_{+}-i\omega)\sqrt{a^{2}b^{2}+8\mu}-\sqrt{2}\left(
aw+2n\right)  bi}{\sqrt{a^{2}b^{2}+8\mu}b}\ ,\\
\beta &  =\frac{(b\delta_{-}-i\omega)\sqrt{a^{2}b^{2}+8\mu}-\sqrt{2}%
(a\omega+2n)bi}{\sqrt{a^{2}b^{2}+8\mu}b}\ ,\\
\gamma &  =1-\frac{2i\omega}{b}\ .
\end{align}

For quasinormal modes we must require that the energy flux at the horizon must
be ingoing, therefore using the fact that $_{2}F_{1}\left(  \mu,\nu
,\sigma,0\right)  =1$ one sees that $C_{2}$ in
(\ref{lasolucion}) must vanish. Consequently, one obtains
\begin{equation}
R\left(  x\right)  =C_{1}\left(  8x+8\mu+a^{2}b^{2}\right)  ^{-i\frac{\sqrt
{2}\left(  a\omega+2n\right)  }{\sqrt{a^{2}b^{2}+8\mu}}}x^{-i\frac{\omega}{b}%
}F\left(  \alpha,\beta,\gamma,-\frac{8x}{a^{2}b^{2}+8\mu}\right)  \ .
\label{laqueentra}%
\end{equation}
Using Kummer relations for the Hypergeometric Functions \cite{AbramowitzStegun}, leads to the following asymptotic behavior at infinity ($x\rightarrow+\infty$) for (\ref{laqueentra}):
\begin{equation}
R\left(  x\right)  \sim\frac{\Gamma\left(  \gamma\right)  \Gamma\left(
\beta-\alpha\right)  }{\Gamma\left(  \gamma-\alpha\right)  \Gamma\left(
\beta\right)  }\frac{1}{x^{\delta_{+}}}\left(  1+\mathcal{O}\left(  \frac
{1}{x}\right)  \right)  +\xi\frac{\Gamma\left(  \gamma\right)  \Gamma\left(
\alpha-\beta\right)  }{\Gamma\left(  \alpha\right)  \Gamma\left(  \gamma
-\beta\right)  }\frac{1}{x^{\delta_{-}}}\left(  1+\mathcal{O}\left(  \frac
{1}{x}\right)  \right)  \ , \label{asymp}%
\end{equation}
where $\xi$ is a non-vanishing constant. In general $\delta_{\pm}=\frac{1}%
{2}\left(  1\pm\sqrt{1-\frac{4\omega^{2}}{b^{2}}}\right)  $ will be complex
numbers, related by $\delta_{+}+\delta_{-}=1$.

The dynamics of the scalar field on the rotating black hole background comes
from the action principle%
\begin{equation}
I\left[  \phi\right]  =-\frac{1}{2}\int\sqrt{-g}d^{3}x\ \nabla_{\nu}%
\Phi\nabla^{\nu}\Phi\ ,
\end{equation}
whose variation reads%
\begin{equation}
\delta I\left[  \phi\right]  =-\int\sqrt{-g}d^{3}x\partial_{\nu}\Phi
\partial^{\nu}\delta\Phi=-\int\sqrt{-g}d^{3}x\left(  \nabla_{\nu}\left(
\nabla^{\nu}\Phi\delta\Phi\right)  -\square\Phi\delta\Phi\right)  \ ,
\end{equation}
which on-shell reduces to a pure boundary term%
\begin{equation}\label{varprin}
\delta_{\text{on-shell}}I\left[  \phi\right]  =-\lim_{r_{0}\rightarrow+\infty
}\left.  \int d^{2}x\sqrt{-g}g^{rr}\partial_{r}\Phi\ \delta\Phi\right\vert
_{r=r_{0}}\ .
\end{equation}
Requiring the action principle to be well-defined implies that $\delta
_{\text{on-shell}}I\left[  \phi\right]  =0$, which using the asymptotic
expansion (\ref{asymp}) implies

\begin{align}
&\left( \frac{  \Gamma\left(  \beta-\alpha\right)  }%
{\Gamma\left(  \gamma-\alpha\right)  \Gamma\left(  \beta\right)  }\right)^{2}%
\frac{ \delta_{+}  }{x^{2\delta_{+}-1}}    +\xi \left( \frac{
\Gamma\left(  \alpha-\beta\right) \Gamma\left(  \beta-\alpha\right)   }{\Gamma\left(  \alpha\right)
\Gamma\left(  \gamma-\beta\right) \Gamma\left(  \gamma-\alpha\right)\Gamma\left(  \beta\right)   } \right)\frac{\left(\delta_{+}+\delta_{-}\right)}{x^{\delta_{+}+\delta_{-}-1}%
}+\nonumber
\\
&\xi^{2}\left(\frac{  \Gamma\left(
\alpha-\beta\right)  }{\Gamma\left(  \alpha\right)  \Gamma\left(  \gamma
-\beta\right)  }\right)^{2}\frac{\delta_{-}}{x^{2\delta_{-}-1}}   =0\
\end{align}

In the limit $x\rightarrow \infty$ the first term goes to zero since $\Re(2\delta_+-1)>0$, the second term is a constant and the third term is divergent since $\Re(2\delta_--1)<0$. The last two terms must vanish independently, and therefore we need to impose that their common factor vanishes by imposing $\Gamma(\alpha)\rightarrow\infty$ or $\Gamma(\gamma-\beta)\rightarrow\infty$. These condition lead to the equations for the spectrum
\begin{equation}
\alpha=-p\ \text{or }\gamma-\beta=-q\ , \label{spec1}%
\end{equation}
where $p$ and $q$ take integer values $0,1,2,3,...\ $.
Conditions (\ref{spec1}) lead to the determination of the spectrum of
quasinormal frequencies for ingoing boundary condition at the horizon and
on-shell vanishing boundary term of the variation of the action functional at infinity (\ref{varprin}). We can introduce the reduced quantities%
\begin{equation}
\omega\rightarrow\sqrt{M}\hat{\omega},\ J\rightarrow\sqrt{M}\hat{J}\ ,
\end{equation}
in terms of which the constraints \eqref{spec1} that determine the spectrum read,%

\begin{align}
\mathcal{C}_1 &  :=-\frac{i\hat{\omega}}{4}-\frac{i\hat{\omega}\hat{J}%
}{2\sqrt{2\hat{J}^{2}+\mu}}-\frac{in}{\sqrt{2\hat{J}^{2}+\mu}}+\frac{1}%
{2}+\frac{\sqrt{4-\hat{\omega}^{2}}}{4}+p=0\ ,\label{constraint1}\\
\mathcal{C}_2 &  :=-\frac{i\hat{\omega}}{4}+\frac{i\hat{\omega}\hat{J}%
}{2\sqrt{2\hat{J}^{2}+\mu}}+\frac{in}{\sqrt{2\hat{J}^{2}+\mu}}+\frac{1}%
{2}+\frac{\sqrt{4-\hat{\omega}^{2}}}{4}+q=0\label{constraint2}\ .
\end{align}
The union of the loci in the complex $\omega$-plane defined by the equations (\ref{constraint1}) and (\ref{constraint2}), defines the quasinormal spectrum of the rotating black hole. The integers $p$ and $q$, determine the overtones. As mentioned above, note that $\mathcal{C}_1(\hat{J},n)=\mathcal{C}_2(-\hat{J},-n)$ with $p$ interchanged by $q$. We observe that the constraint $\mathcal{C}_1=0$($\mathcal{C}_2=0$) can be solved only for positive(negative) values of the black hole angular momentum $\hat{J}$, but due to the symmetry mentioned above, as expected, both constraints will lie on the same locus.

Consequently, in terms of the reduced angular momentum $\hat{J}=J/\sqrt{M}$, the reduced frequencies $\hat{\omega}=\omega/\sqrt{M}$ can be conveniently summarized as
\begin{equation}
\hat{\omega}=-\frac{\left(  2n\sigma_{\hat{J}}+i\left(  1+2p\right)  \nu\right)  }%
{2|\hat{J}|\left(  |\hat{J}|+\nu\right)  }\left(  2|\hat{J}|+\nu+i\nu 
\sigma_{n \cdot \hat{J}}\sqrt{\frac{4|\hat{J}|\left(  
|\hat{J}|+\nu\right)
}{\left(  2\sigma_{\hat{J}}n+i\left(  1+2p\right)  \nu\right)  ^{2}}%
-1}\right) \,
\end{equation}
where we have defined $\nu:=\sqrt{2\hat{J}^{2}+\mu}$ and set Newton's constant $G=1/2$. $\sigma_x$ stands for the sign of $x$, defined such that $\sigma_0=1$.
\\
Some comments on the boundary conditions are in order. For asymptotically flat
black holes in four dimensions, as for example in the Kerr family, the scalar
field has two possible behaviors at infinity which are plane waves that
describe the out-going and the in-coming modes, and therefore the leading
contribution goes as $e^{-i\omega\left(  t\pm r\right)  }$ with the plus(minus)
sign standing for the in-coming(out-going) mode. The background of the rotating
asymptotically, locally flat, black hole of New Massive Gravity allows different
behaviors at infinity, which are given in (\ref{asymp}). Even in terms of the
proper radial distance at infinity $\rho=\frac{2}{\sqrt{b}}\sqrt{r}$, one would
have $x\sim\rho^{2}$, and the behavior (\ref{asymp}) would still have a
power law fashion and would not be compatible with the interpretation of an
out-going or in-coming plane wave. The situation at hand is actually more
similar to the analysis of quasinormal modes in asymptotically AdS black holes
where the leading asymptotic behaviors are $r^{-\Delta_{\pm}}$ and $\Delta_{\pm}$ are
given in terms of the mass of the field and the dimension of the spacetime, defining the conformal weights of the operator dual to the scalar on the
boundary CFT. Therefore, with the asymptotic behaviors given by equation (\ref{asymp}) a natural
physical condition that allows to compute the quasinormal frequencies is to
require that the field at infinity decays fast enough as to lead to a well
defined variational principle (\ref{varprin}) and this is in fact the strategy we have followed.

Below we present plots of the QNM spectrum of the scalar probe in terms of the physical variables, as well as closed analytic expressions for the asymptotic behavior of the frequencies.

\subsection{Static case and asymptotic frequencies}

Let's analyze first the static case. We set $J=0$ in (\ref{constraint1}) and (\ref{constraint2}). Then from
(\ref{constraint1}) one obtains two equations for the real and imaginary part of the constraint, whose real part reads%
\begin{equation}
\frac{1}{2}+p+\frac{1}{2\sqrt{2}}\left(  \omega_{i}+\left(  4\omega_{i}%
^{2}\omega_{r}^{2}+\left(  2+\omega_{i}^{2}-\omega_{r}^{2}\right)
^{2}\right)  ^{1/4}\cos\left(  \frac{\arg\left(  2-\left(  \omega_{r}%
+i\omega_{i}\right)  ^{2}\right)  }{2}\right)  \right)  =0\ .
\label{constraintsinsolucion}%
\end{equation}
This equation does not have solution for any value of
$\omega=\omega_{r}+i\omega_{i}$ (with $\omega_{r,i}$ finite), and therefore (\ref{constraint1}) cannot be fulfilled. A
similar analysis of the real part of (\ref{constraint2}) leads to the same equation (with $p$ changed by $q$). This is a peculiar feature of
the static solution: there are no massless scalar field quasinormal modes for
the asymptotically flat, static black hole with gravitational hair in NMG. The constraints (\ref{constraint1}) and (\ref{constraint2}) can be
mapped to the Riemann sphere, where it can be seen that at the leading order
when $\omega_{i}\rightarrow-\infty$ (for constant $\omega_{r}$) the constraint
is indeed solved since (\ref{constraintsinsolucion}) reads%
\begin{equation}
\omega_{i}+|\omega_{i}|\cos\left(  \frac{\arg(\omega_{i}^{2})}{2}\right)
+...=0\ ,
\end{equation}
which vanishes when $\omega_{i}\rightarrow-\infty$. This result is recovered from
the quasinormal modes of the rotating case. We will see that as $J\rightarrow 0$, the imaginary part of the quasinormal frequencies
tend to $-\infty$, and therefore the quasinormal modes acquire an infinity
damping in this limit, and migrate to the north pole of the Riemann Sphere\footnote{Same conclusion is obtained if one starts with the asymptotically locally flat, static black hole from the scratch. Quasinormal modes of scalar and Dirac fields on the asymptotically AdS, hairy black hole of NMG have been explored numerically and perturbatively in \cite{Kwon:2011ey} and \cite{Gonzalez:2014voa}.}. 

\subsection{The quasinormal spectrum}
In this section we present different plots that clarify the behavior of the spectra as a function of the parameters of the problem, with descriptive captions. 
\begin{figure}[h!]
\centering
\includegraphics[scale=0.2]{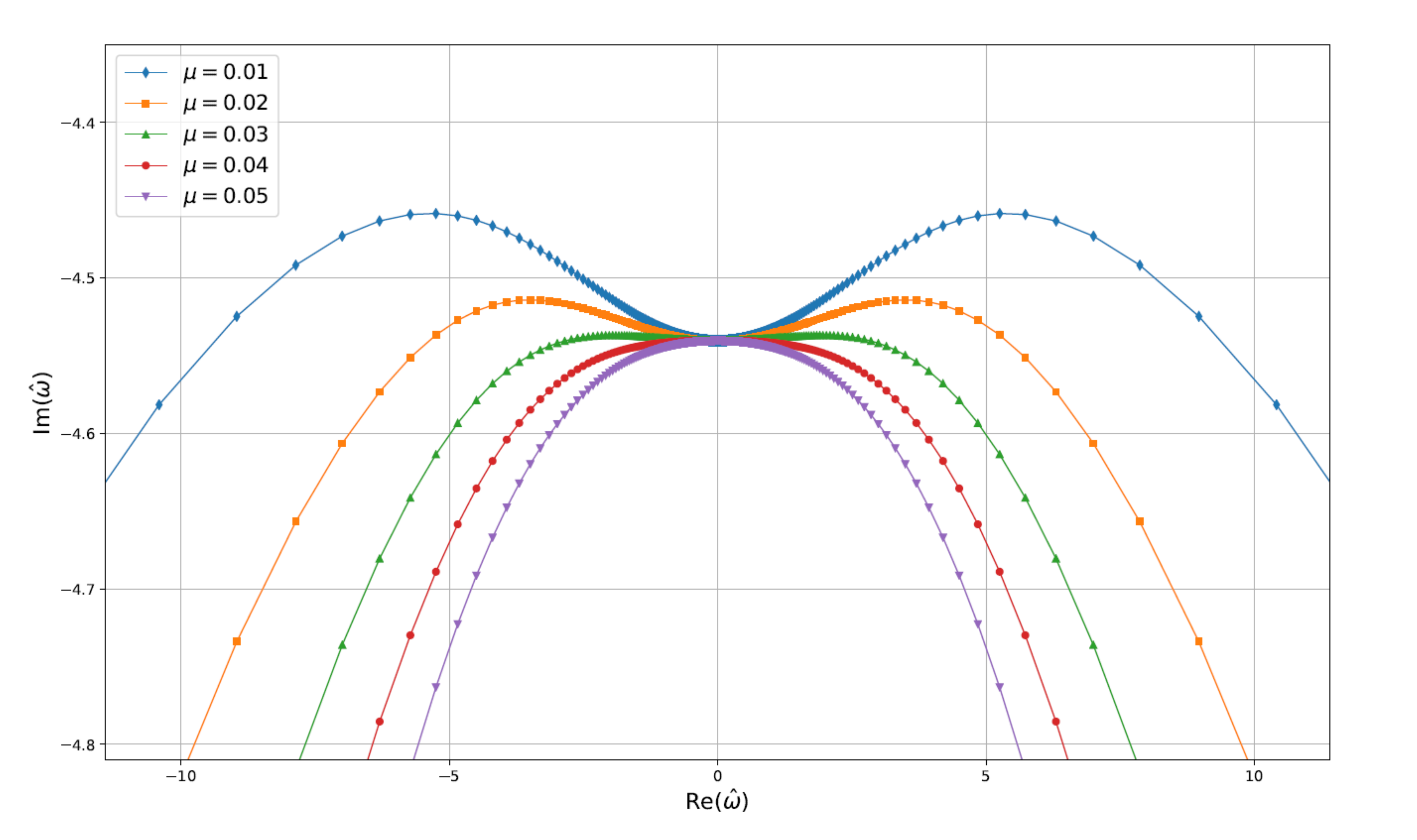}\qquad\includegraphics[scale=0.25]{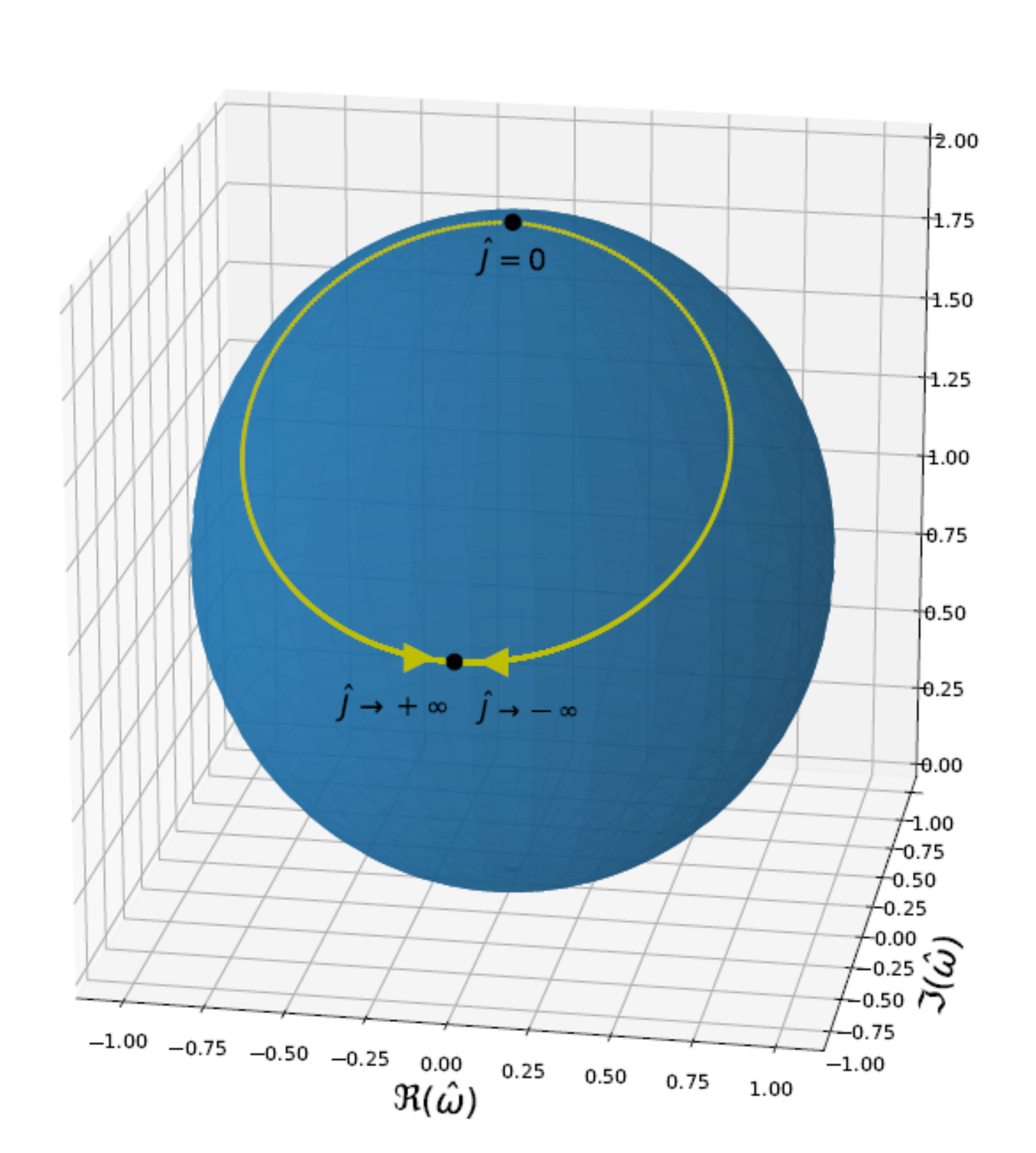}
\caption{\small{Left panel: Spectra for different values $\mu$, and $p=1$. For $n=1$. The curves start on the imaginary axis ($\hat{J}\rightarrow-\infty$), then acquire a positive real part, approach the north pole of the Riemann Sphere ($\hat{J}\rightarrow0^-$), then pass through the pole and acquire a negative real part, and end again in the imaginary axis as $\hat{J}\rightarrow+\infty$. For $n=-1$ the locus is exactly the same but has the opposite orientation. The right panel shows the projection to the Riemann sphere of the spectrum for $\mu=0.1$, $p=0$ and $n=1$ showing that the north pole is reached as $\hat{J}$ goes to zero.}}%
\end{figure}

\begin{figure}[h!]
\centering
\includegraphics[scale=0.23]{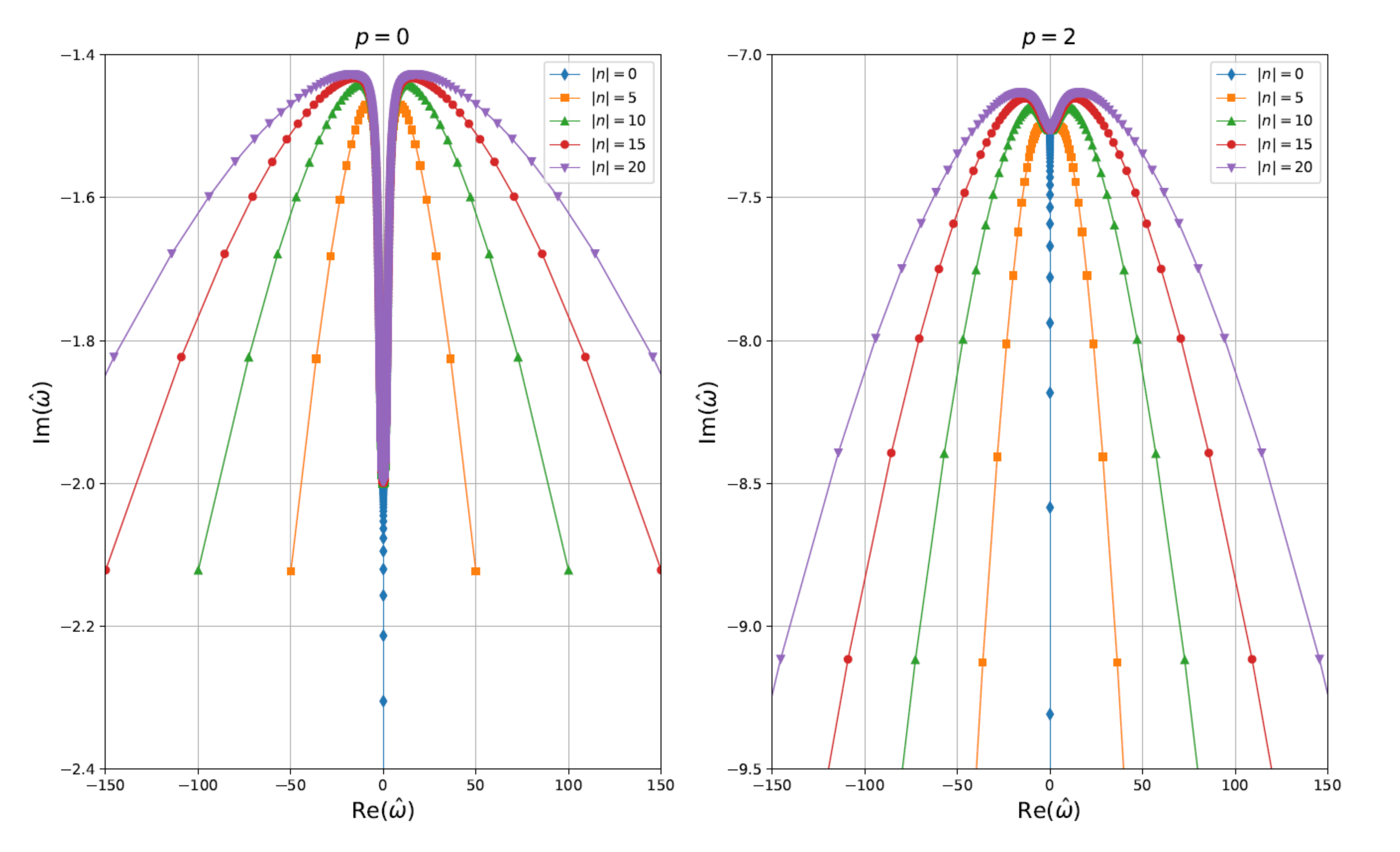}
\caption{\small{Spectrum for different values of $n$. The modes with $n=0$ lie on the imaginary axis for any $J$ which runs along the curves. We have fixed $\mu=0.01$ and left panel shows the fundamental mode $p=0$ while right panel shows the modes with $p=2$. As expected, for a given value of the reduced angular momentum $\hat{J}$, the modes with higher overtones have a higher damping.}}%
\end{figure}

\begin{figure}[h!]
\centering
\includegraphics[scale=0.3]{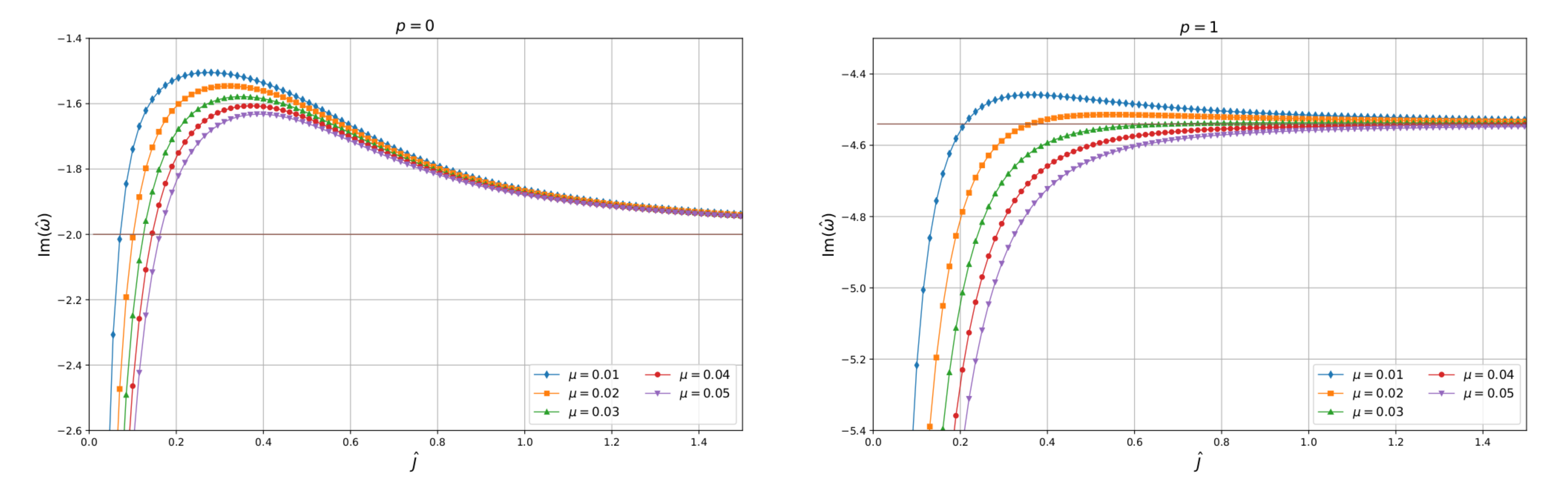}
\caption{\small{Imaginary part of the frequency (always negative) as a function of the reduced angular momentum $\hat{J}$ for different values of $\mu$, and $n=1$. For large angular momentum the imaginary part of the frequencies approaches a finite constant ($p=0$ left panel and $p=1$ right panel).}}%
\end{figure}

\begin{figure}[h!]
\centering
\includegraphics[scale=0.25]{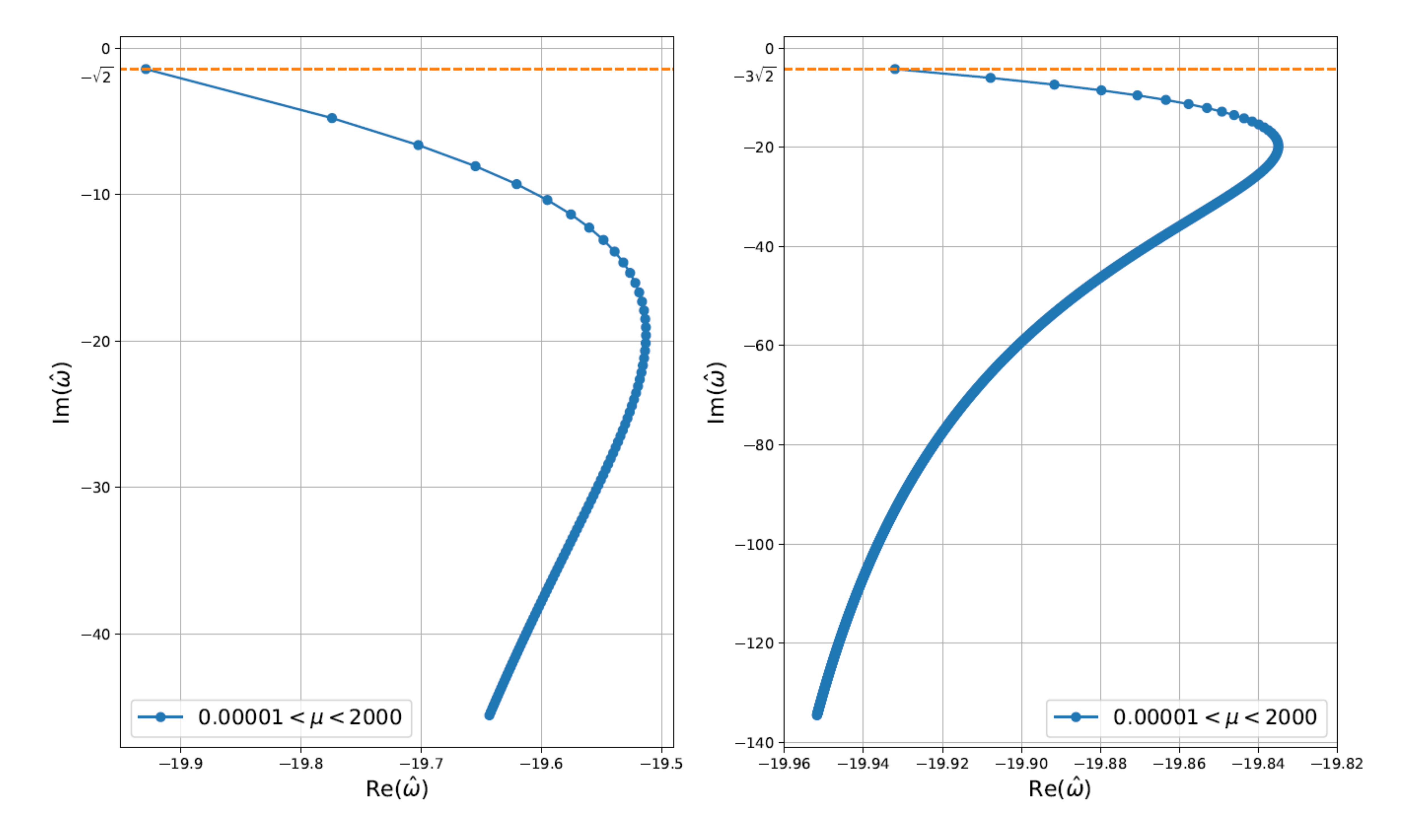}
\caption{\small{Quasinormal spectrum for the fundamental ($p=0$) and first excited mode ($p=1$) for different values of the gravitational hair parameter $\mu$. As $\mu$ increases the modes acquire a higher damping (we have set $\hat{J}=n=1$). We observe that $\Im(\hat\omega)<-\sqrt{2}$ and approaches the bound for small $\mu$ and large $n$ and $p=0$. For modes with higher overtones the negative, upper bound on the imaginary part of the frequencies is always lower, leading to higher damping.}}%
\end{figure}

\begin{figure}[h!]
\centering
\includegraphics[scale=0.25]{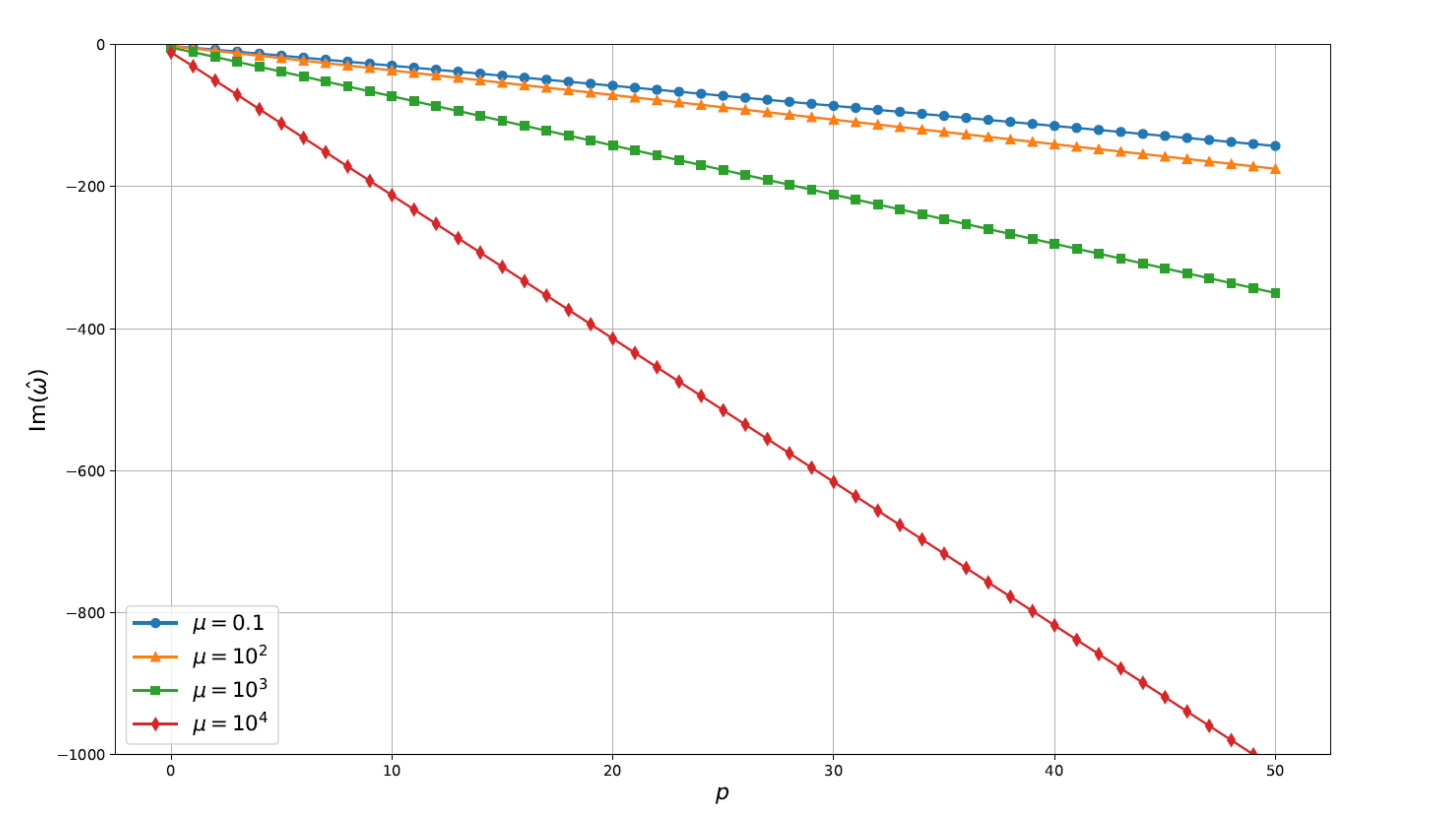}
\caption{\small{Imaginary part of the frequencies as a function of the mode number $p$ for different values of the gravitational hair parameter $\mu$. As the latter increases the modes acquire a higher damping. We have chosen $\hat{J}=10$ and $n=1$.}}%
\end{figure}

\newpage
\subsection{Asymptotic frequencies}
Finally let's analyze two informative asymptotic behaviors for the spectrum. First, for large angular momentum of the field the asymptotic expansion $n\rightarrow+\infty$ reads%
\begin{equation}\label{spectrumlargen}
\hat{\omega}_{n\rightarrow+\infty}=-\frac{2n}{\hat{J}}-i(1+ 2 p)\sqrt{2+\frac{\mu}{\hat{J}^{2}}}+\left(\mathcal{O}\left(\frac{1}{n}\right)+\mathcal{O}\left(\frac{1}{n^2}\right)i\right)\ .
\end{equation}
It is interesting to notice that this behavior is reminiscent of that
of the quasinormal modes of a scalar field with large angular momentum $j$ on Schwarzschild black hole. In that case one has \cite{Ferrari:1984zz}%
\begin{equation}
\omega\rightarrow\pm\frac{1}{3\sqrt{3}M}\left(  j+\frac{1}{2}\right)
-\frac{1}{6\sqrt{3}M}\left(  1+2p\right)  i+\mathcal{O}\left(  \frac{1}%
{j}\right)  \ .
\end{equation}

It is also interesting to put special attention to the case with arbitrarily large $J$, since this cannot be reached within the Kerr family. (It can be reached in Myers-Perry in $d>5$ \cite{Myers:1986un}). Quasinormal frequencies $\hat{\omega}$ in the limit $\hat{J}\rightarrow \infty$ becomes purely imaginary (damped) and depend only on the overtone integer
\begin{equation}
  \hat{\omega}_{\hat{J}\rightarrow \infty} = - i \left(1+2p+\sqrt{1+4\left(\sqrt{2}-1\right)^2p\left(1+p\right) }\right)\,\,.
\end{equation}
The fundamental mode in this regime is $  \hat{\omega}_{\hat{J}\rightarrow \infty}^{0}=-2\, i$. The function $\hat{\omega}_{\hat{J}\rightarrow \infty}(p)$ at large $p$ approaches a straight line with slope $2\sqrt{2}$.

\section{Conclusions}

In this paper we have shown that a massless scalar probe on the asymptotically locally flat, rotating black hole of pure NMG can be analytically solved leading to an exact expression for the quasinormal mode frequencies. We have imposed ingoing boundary conditions at the horizon and a decay at infinity which is fast enough leading to a well defined variational principle for the scalar probe. For a given black hole, parametrized by the mass, angular momentum and gravitational hair, the quasinormal spectrum is defined by the angular momentum of the field as well as an integer counting overtones. We showed that under these boundary conditions the imaginary part of the frequencies is always negative, not leading to any superradiance, as expected for a massless scalar. Even more, when the angular momentum of the black hole approaches zero, the imaginary part of the frequencies tends to minus infinity leading to infinitely damped modes. Remarkably, this is one of the few examples of rotating black holes in which the quasinormal frequencies for a scalar probe can be obtained in an analytic, closed form\footnote{The other examples being BTZ \citep{ExactBTZ}, the so-called subtracted geometries \citep{subtracted} and near horizon geometries of rotating black holes \citep{nh}. For a thorough review on QNMs see \citep{Berti:2009kk}.}.

Since the rotating black hole, even in the case with cosmological constant, has vanishing Cotton tensor \citep{OTT}, \citep{Bergshoeff:2009aq}, it is a conformally flat spacetime, therefore  the solution for a conformal scalar could also be obtained in an analytic manner by applying a conformal transformation to the solution of the free, massless scalar of 3D Minkowski or AdS space (since the latter is also conformally flat). Once the mapping is done on the general solution, one would have to impose the boundary conditions at the horizon and infinity taking care of the potential divergences induced by the conformal mapping that leads to the black hole.

The stable propagation of scalar probes on the rotating background may lead to the existence of rotating black holes with both scalar and gravitation hair. In NMG static black holes with self-interacting scalar hair already exist, with Lifshitz and AdS asymptotics \citep{Correa:2014ika}, \cite{Ayon-Beato:2015jga}, and it would be interesting to construct rotating black holes with these hairs.
\bigskip

\section*{Aknowledgements}
The authors want to thank Gaston Giribet and Nicolas Grandi for enlightening comments. This work was partially supported by FONDECYT grants 1170279 and 1181047. O.F. is supported by the Project DINREG 19/2018 of the Dirección de Investigación of the Universidad Católica de la Santísima Concepción.

\end{document}